\documentclass[journal]{IEEEtran}

\usepackage{xcolor,soul,framed} %,caption
\usepackage[pdftex]{graphicx}
\graphicspath{{../pdf/}{../jpeg/}}
\DeclareGraphicsExtensions{.pdf,.jpeg,.png}
\usepackage[cmex10]{amsmath}
\usepackage{times}
\usepackage{url}
\usepackage[hidelinks]{hyperref}
\usepackage[utf8]{inputenc}
\usepackage[small]{caption}
\usepackage{amsthm}
\usepackage{booktabs}
\usepackage{algorithm}
\usepackage{algorithmic}
\usepackage[switch]{lineno}
\usepackage{amsfonts}
\usepackage{multirow}
\usepackage{subfigure}
\usepackage{bbding}
\usepackage{dsfont}
\usepackage{bm}
\usepackage{colortbl}
\usepackage{amssymb}
\usepackage[normalem]{ulem}

\newtheorem{theorem}{Theorem}

\begin{document}
	\bstctlcite{IEEEexample:BSTcontrol}
	\title{Fair Federated Medical Image Classification Against Quality Shift via Inter-Client Progressive State Matching}
	\author{Nannan Wu,
		Zhuo Kuang,
		Zengqiang Yan,
		Ping Wang,~\IEEEmembership{Fellow,~IEEE,}
		and Li Yu,~\IEEEmembership{Senior Member,~IEEE,}% <-this % stops a space
		\thanks{Nannan Wu, Zhuo Kuang, Zengqiang Yan, and Li Yu are with the School of Electronic Information and Communications, Huazhong University of Science and Technology, Wuhan, China (e-mail: wnn2000@hust.edu.cn; kuangzhuo@hust.edu.cn; z\_yan@hust.edu.cn; hustlyu@hust.edu.cn)}% <-this % stops a space
		\thanks{Ping Wang is with the Department of Electrical Engineering and Computer Science, Lassonde School of Engineering, York University, Toronto, Canada (e-mail:  ping.wang@lassonde.yorku.ca).}
		\thanks{This is the preprint version.}
	}

	\markboth{Preprint}{}

	% ====================================================================
	\maketitle

	% === ABSTRACT ====================================================================
	% =================================================================================
	\begin{abstract}
		%\boldmath
		
		Despite the potential of federated learning in medical applications, inconsistent imaging quality across institutions—stemming from lower-quality data from a minority of clients—biases federated models toward more common high-quality images. This raises significant fairness concerns. Existing fair federated learning methods have demonstrated some effectiveness in solving this problem by aligning a single 0th- or 1st-order state of convergence (\textit{e.g.}, training loss or sharpness). However, we argue in this work that fairness based on such a single state is still not an adequate surrogate for fairness during testing, as these single metrics fail to fully capture the convergence characteristics, making them suboptimal for guiding fair learning. To address this limitation, we develop a generalized framework. Specifically, we propose assessing convergence using multiple states, defined as sharpness or perturbed loss computed at varying search distances. Building on this comprehensive assessment, we propose promoting fairness for these states across clients to achieve our ultimate fairness objective. This is accomplished through the proposed method, FedISM+. In FedISM+, the search distance evolves over time, progressively focusing on different states. We then incorporate two components in local training and global aggregation to ensure cross-client fairness for each state. This gradually makes convergence equitable for all states, thereby improving fairness during testing. Our empirical evaluations, performed on the well-known RSNA ICH and ISIC 2019 datasets, demonstrate the superiority of FedISM+ over existing state-of-the-art methods for fair federated learning. The code is available at \textcolor{blue}{https://github.com/wnn2000/FFL4MIA}.

	\end{abstract}

	\begin{IEEEkeywords}
		Federated Learning, Fairness, Medical Image Classification, State Matching
	\end{IEEEkeywords}

	% \fi
	\IEEEpeerreviewmaketitle

	\section{Introduction}
	
	\IEEEPARstart{I}{n} response to growing concerns over data privacy, federated learning (FL) \cite{mcmahan2017communication} has been recognized as a promising paradigm for safeguarding sensitive information through decentralized training of deep neural networks, particularly in medical domains \cite{dou2021federated}. Despite its potential, FL faces a significant challenge of data heterogeneity across medical institutions \cite{ye2023heterogeneous,FCCL_CVPR22,FCCLPlus_TPAMI23}, resulting from independent data acquisition processes led by different institutions. Existing research has examined this heterogeneity from various perspectives, including domain shift \cite{li2021fedbn,liu2021feddg,jiang2023iop}, label distribution skew \cite{zhang2022federated,DBLP:conf/miccai/WuYYCY23}, and label quality variation \cite{DBLP:conf/ijcai/Wu0JCY23,DBLP:journals/mia/ChenLXY23,wu2023feda3i,wicaksana2023fedmix}. Nevertheless, the prevalent issue of quality heterogeneity \cite{fang2023robust} in medical imaging, which may pose new challenges for FL, has not been fully explored.
	
	\begin{figure}[!t] 
		\centering
		\includegraphics[width=1.0\columnwidth]{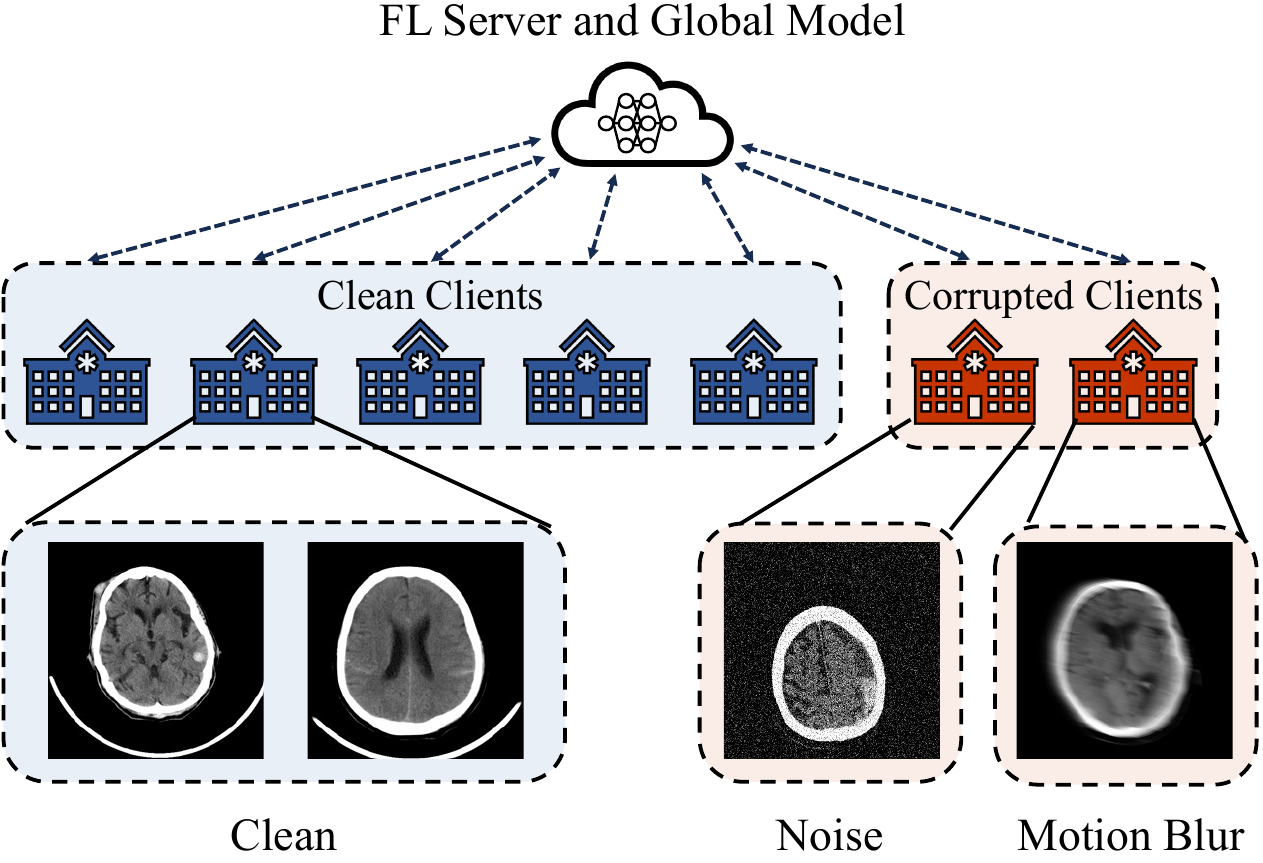}
		\caption{Imaging quality shifts across clients, where most clients have high-quality (\textit{i.e.}, clean) images, while others experience corruption, such as noise or motion blur.}
		\label{fig:background}
	\end{figure}
	
	Although strict protocols are followed in medical imaging environments, image quality uniformity across institutions is not guaranteed. As depicted in Fig.~\ref{fig:background}, variations in equipment and imaging conditions lead to discrepancies in image quality, resulting in corruption in some images—such as noise from equipment malfunctions or motion blur from involuntary movements—while others remain unaffected. Typically, only a small proportion of images are low-quality (\textit{i.e.}, corrupted) \cite{DBLP:conf/cvpr/HuangZX00000L23}, creating a special quality shift. This shift causes standard FL methods, \textit{e.g.}, FedAvg \cite{mcmahan2017communication}, to prioritize the more common clean images, leading to degraded performance on the rarer corrupted data and limiting the applicability of federated models. In this work, we pioneer the identification of this real-world challenge as the significant performance disparity across data quality and demonstrate theoretically that addressing this issue is equivalent to ensuring Rawlsian Max-Min fairness \cite{rawls2001justice} for client performance \cite{FedISM} (see Sec. \ref{sec:setting}). This issue is then framed as a client-level fairness problem, namely: \textbf{\textit{how can we enhance performance lower bound across clients with varying image quality distributions, including both clean and corrupted images?}} Our study marks the inaugural effort to address fairness in FL across clients experiencing imaging quality shifts, which differs from existing studies that focus on fairness under domain \cite{jiang2023fair} and label distribution shifts \cite{li2019fair}.

	\begin{figure}[!t] 
		\centering
		\includegraphics[width=1.0\columnwidth]{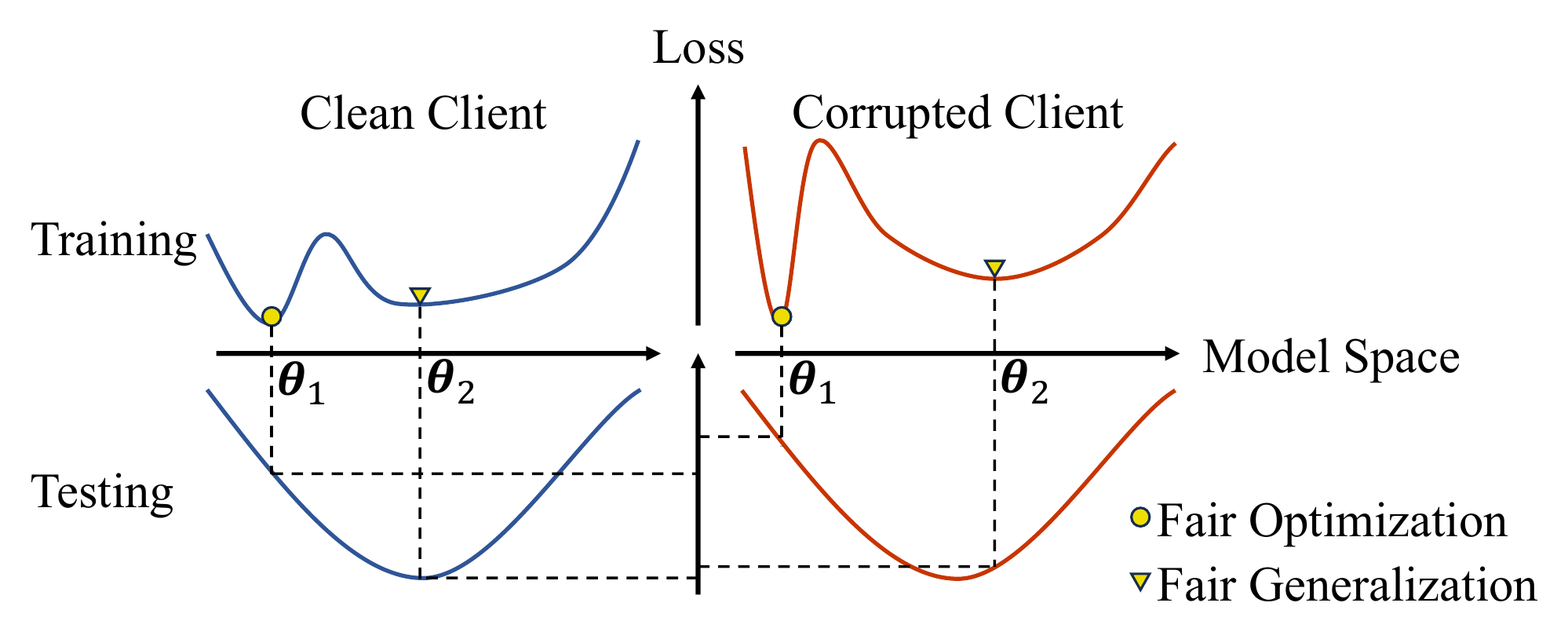}
		\caption{Comparison of fair optimization and fair generalization: Fair optimization typically leads to convergence at sharp minima (with model parameters $\boldsymbol{\theta}_1$), resulting in uniformly low loss values but poor testing performance. In contrast, fair generalization emphasizes uniformly low sharpness (with model parameters $\boldsymbol{\theta}_2$), which improves absolute performance and fairness when testing.}
		\label{fig:motivation}
	\end{figure}

	Previous fair FL solutions can be categorized into two approaches. The first, summarized as \textbf{\textit{fair optimization}} (see Sec. \ref{sec:fair_op}), focuses on improving fairness during the optimization process across different clients/distributions \cite{jiang2023fair,li2019fair,mohri2019agnostic,ezzeldin2023fairfed,zhang2023federated}. Specifically, they monitor a client-specific indicator (\textit{e.g.}, training loss), which directly reflects the degree of the optimization process carried out on each distribution. In response to the selected indicator, the methods dynamically adjust the weights assigned to clients in FL, prioritizing those with poorer performance. This strategy can ultimately reduce the upper bound of the loss across all clients, backed by theoretical guarantees \cite{li2019fair}, thereby promoting fairness.
	
	However, this approach is sub-optimal as it only considers the 0th-order state of convergence\footnote{The 0th-order state of convergence refers to a specific metric (\textit{e.g.}, loss or accuracy) evaluated at convergence itself.}, which captures optimization on training data but fails to fully assess generalization to testing data. As shown in Fig. \ref{fig:motivation}, strict adherence to this metric may lead to superficial fairness, such as convergence to a sharp minimum for all clients that severely impairs generalization \cite{foret2021sharpness,zhuang2022surrogate}. Minimizing such a gap points to the second approach, \textit{i.e.}, \textbf{\textit{fair generalization}} (see Sec. \ref{sec:fair_generalization}). Motivated by recent advances in sharpness-aware minimization, which highlight the inverse correlation between generalization performance and the 1st-order state of convergence\footnote{The 1st-order state of convergence  refers to the rate of change or trend of a specific metric (\textit{e.g.}, loss or accuracy) near convergence} (\textit{e.g.}, sharpness) \cite{foret2021sharpness,zhuang2022surrogate}, a preliminary work proposes to minimize and uniformize sharpness across clients \cite{FedISM}, as shown in Fig. \ref{fig:motivation}. This is accomplished through a method, federated learning via inter-client sharpness matching (FedISM) \cite{FedISM}, which computes sharpness empirically within a predetermined search distance.
	
	Though such a 1st-order state (\textit{i.e.}, sharpness) is better than the 0th-order state for guiding fair learning, it still fails to fully capture the comprehensive characteristics of convergence on the loss surface. For instance, a convergence with uniformly low sharpness (at a fixed search distance) on different distributions can still exhibit varying high loss and high sharpness (at other search distances). This phenomenon leads to sub-optimal generalization, because the optimal sharpness, which needs to be minimized for better performance, is defined at different search distances across distributions \cite{DBLP:conf/cvpr/HuangZX00000L23}. This indicates that a single state, whether 0th- or 1st-order, is not a reliable surrogate for assessing convergence and guiding the process of promoting fairness. We therefore focus on the following question: \textit{\textbf{What is the precise surrogate for the generalization ability of the convergence?}}
	To address this, we propose a comprehensive framework by considering multiple states (see Sec. \ref{sec:assessment}). Specifically, we compute sharpness or perturbed loss at varying search distances, from zero to a maximum. This enables us to pursue the final fairness target by achieving fairness for all these states of convergence across clients. We note that this new framework generalizes previous approaches—fair optimization and fair generalization—where the search distance is fixed at 0 or a positive constant, respectively. Thus, our framework provides a better assessment for convergence.
	
	Based on such comprehensive assessment of convergence, the key to the final fairness target is \textit{\textbf{how to achieve fairness for these states across clients.}}
	To address this, we propose an enhanced FL method called inter-client progressive states matching (\textbf{FedISM+}) (see Sec. \ref{sec:FedISM+}). We first introduce a scheme that gradually adjusts the search distance during training, allowing us to focus on different states at different stages of FL. For each state, FedISM+ incorporates sharpness-aware local optimization, with each client minimizing sharpness. During global aggregation, weights are adjusted based on each client’s sharpness or perturbed loss level. This ensures that the global update effectively reduces sharpness/perturbed loss, particularly for clients with higher initial levels, resulting in a more uniform state distribution across clients. Throughout the entire FL process, FedISM+ progressively focuses on different states, considering fairness for all states.
	
	We note that FedISM+ is not only easy to implement but also highly effective. Compared to FedAvg \cite{mcmahan2017communication}, it requires only three lines of code modifications. Its effectiveness has been validated through extensive empirical evaluations on two widely used medical image datasets, demonstrating superior performance over state-of-the-art methods (see Sec. \ref{sec:eval}). We also discuss its communication and privacy properties (see Sec. \ref{sec:privacy}).
	
	This paper is a \textbf{substantial extension of our preliminary work} \cite{FedISM} reported in IJCAI-2024 from the following main aspects:
	
	\begin{itemize}
		\item New Insight: Using a single state to assess convergence and guide the process of promoting fairness, as done in \cite{FedISM}, is inadequate for either 1st-order or 0th-order states.
		
		\item Comprehensive Convergence Assessment: We introduce a new framework that generalizes previous methods relying on a single state for convergence assessment and fair learning, by considering multiple states.
		
		\item An Enhanced FL Method: Based on this comprehensive assessment, we propose achieving the final fairness goal by ensuring cross-client fairness for all these states of convergence. This is done by progressively aligning sharpness/perturbed loss at different search distances, from zero to a maximum. By considering multiple states, this approach effectively combines fair optimization and fair generalization.
		
		\item Extensive Empirical Evaluation: We assess the performance of FedISM+ by conducting extensive experiments on two real-world medical image classification datasets, \textit{i.e.}, RSNA ICH and ISIC 2019, demonstrating its superiority over current FL methods in enhancing fairness despite variations in imaging quality.
		
	\end{itemize}
	
	\section{Related Work}
	\subsection{Fair Federated Learning}
	Fairness \cite{huang2023federated} has emerged as a vital topic in FL, focusing mainly on collaborative fairness \cite{lyu2020collaborative,xu2021gradient} and performance fairness \cite{li2019fair}. Collaborative fairness ensures that each participant’s reward is proportional to their contribution to the federation, which is crucial for maintaining motivation and engagement among clients. In contrast, this paper focuses on performance fairness, which typically strives for uniformly high performance (\textit{i.e.}, a low upper bound of loss as Eq. \ref{eq:client_fairness}) across various devices/features/distributions, thereby ensuring equal benefits from the federation. Current solutions primarily aim to achieve this goal by aligning the single state of convergence across clients. A series of works focus on the 0th-order state \cite{jiang2023fair,li2019fair,mohri2019agnostic,ezzeldin2023fairfed,zhang2023federated}, while our preliminary work in \cite{FedISM} focuses on the 1st-order state \cite{FedISM}. However, these approaches fail to assess convergence comprehensively, leading to a significant gap between the fairness measured by the chosen metric and the final fairness target.
	
	\subsection{Sharpness of Loss Surface}
	A major challenge in machine learning is to reconcile training optimization with testing generalization. Recent research on sharpness-aware minimization suggests that models generalize more effectively when they converge to flat minima, as opposed to sharp ones \cite{foret2021sharpness,zhuang2022surrogate}. Building on this understanding, several studies have integrated sharpness metrics to address issues of poor generalization. For instance, SharpDRO \cite{DBLP:conf/cvpr/HuangZX00000L23} combines sharpness with GroupDRO \cite{sagawa2019distributionally} to enhance robust generalization. ImbSAM \cite{zhou2023imbsam} targets enhancing recognition of tail classes in long-tailed image datasets by reducing the sharpness within these classes. While these approaches have been extensively studied in traditional machine learning contexts, their application in FL has gained attention \cite{qu2022generalized,caldarola2022improving,sun2023dynamic}. Yet, the specific relationship between sharpness and fairness within FL remains largely unexplored.

	\section{Methodology}
	
	\subsection{Preliminaries}  \label{sec:setting}
	For a standard image classification task with $C$ classes, we consider a cross-silo FL setting involving $K$ participants. Each participant $k$ holds a private dataset $D_k=\{(\boldsymbol x_i \in \mathcal{X}, y_i \in \mathcal{Y})\}_{i=1}^{N_k}, k \in [K] = \{1,2,\cdots, K\}$, where $\mathcal{X}$ and $\mathcal{Y}=[C] = \{1, 2, \cdots, C\}$ represent the input image and label spaces, respectively. Each image-label pair $(\boldsymbol x_i, y_i)$ is drawn from the client-specific distribution $\mathbb{P}_k(\boldsymbol x,y \mid a_k)$, with $a_k$ indicating an attribute influencing the image quality in client $k$, \textit{i.e.}, $\mathbb{P}_k(\boldsymbol x,y \mid a_k)=\mathbb{P}_k(\boldsymbol x \mid y, a_k)\mathbb{P}_k(y)$. This paper highlights that this quality attribute is non-identical across clients, \textit{i.e,} $\{a_1, a_2, \cdots, a_K\} = [A]$ and $A > 1$.
	Let $f(\cdot; \boldsymbol \theta): \mathcal{X} \rightarrow \Delta^{C-1}$ represent a deep neural network parameterized by $\boldsymbol \theta$, and $\ell: \Delta^{C-1} \times \mathcal{Y} \rightarrow \{0\} \cup \mathbb{R+}$ denote the loss function, where $\Delta$ is the probability simplex. The goal of this paper is to train a model that achieves Rawlsian Max-Min fairness \cite{rawls2001justice} with respect to imaging quality, formulated as:
	\begin{equation} \label{eq:objective}
		\boldsymbol \theta^* = \mathop{\arg\min}\limits_{\boldsymbol \theta} \{\max_{a \in [A]} \mathbb{E}_{(\boldsymbol x,y) \sim \mathbb{P}(\boldsymbol x, y \mid a)} [\ell(f(\boldsymbol x; \boldsymbol \theta), y)] \}.
	\end{equation}
	
	Fair learning that prioritizes the worst group is required to achieve Eq. \ref{eq:objective}.
	In traditional centralized learning with all data aggregated, this can be effectively solved through distributionally robust optimization \cite{sagawa2019distributionally}, leveraging attribute information (\textit{i.e.}, the imaging quality of each image) obtained in advance. However, such prior knowledge is unavailable due to privacy constraints in FL, making attribute-aware design impractical. Therefore, we approach the problem in a data-agnostic manner inspired by \cite{papadaki2022minimax}, as outlined in the following theorem:
	\begin{theorem}[Equivalence] \label{th:Equivalence}
		Assuming label distributions of all clients are identical, we have:
		\begin{equation} \label{eq:client_fairness}
			\resizebox{0.9\hsize}{!}{$
				\begin{aligned}
					&\boldsymbol \theta^*, \boldsymbol \lambda^* = \arg \mathop{\min}\limits_{\boldsymbol \theta} \mathop{\max}\limits_{\boldsymbol \lambda \in \Delta^{K-1}} \sum_{k=1}^{K} \lambda_k\mathbb{E}_{(\boldsymbol x,y) \sim \mathbb{P}_k(\boldsymbol x, y \mid a_k)} [\ell(f(\boldsymbol x; \boldsymbol \theta), y)], 
				\end{aligned}
				$}
		\end{equation}
		and
		\begin{equation} \label{eq:group_fairness}
			\resizebox{0.9\hsize}{!}{$
				\begin{aligned}
					&\boldsymbol \theta^*, \boldsymbol \mu^* = \arg \mathop{\min}\limits_{\boldsymbol \theta} \mathop{\max}\limits_{\boldsymbol \mu \in \Delta^{A-1}} \sum_{u=1}^{A} \mu_u \mathbb{E}_{(\boldsymbol x,y) \sim \mathbb{P}(\boldsymbol x, y \mid u)} [\ell(f(\boldsymbol x; \boldsymbol \theta), y)],
				\end{aligned}
				$}
		\end{equation}
		where $\boldsymbol \lambda$ and $\boldsymbol \mu$ denote client- and attribute-wise weights, respectively, and $\mu^*_u = \sum_{k=1}^{K} \mathds{1}_{a_k=u} \lambda^*_k$.
	\end{theorem}
	\noindent As shown above, prioritizing low-performance clients with high expected risk (Eq. \ref{eq:client_fairness}), \textit{i.e.}, promoting Rawlsian Max-Min fairness \cite{rawls2001justice} for client performance, inherently satisfies the primary objective (Eq. \ref{eq:group_fairness}). It is important to note that even in the presence of heterogeneity in label distributions, this conclusion remains unchanged, as such heterogeneity can be effectively addressed using logit adjustment techniques \cite{menon2020long, zhang2022federated}. In summary, Theorem \ref{th:Equivalence} provides a solution to the problem that bypasses the quality attribute. We therefore shift the focus of this paper to the newly constructed objective in Eq. \ref{eq:client_fairness}, \textit{i.e.}, how to achieve performance-fair FL across clients with varying image quality.
	
	\subsection{Previous Solutions: Single State Alignment} \label{sec:previous_solutions}
	Client performance unfairness in FL often arises from the neglect of certain clients with scarce data or atypical distributions (\textit{i.e.}, clients with corrupted images in this work), as improving their performance has negligible impact on the overall objective. To address this issue, previous solutions propose achieving Rawlsian Max-Min fairness \cite{rawls2001justice} in specific metrics, which can be categorized into two main approaches:
	
	\subsubsection{Fair Optimization}\label{sec:fair_op}
	The training loss reflects how well the model fits the training data, making fairness in this metric a potential surrogate for performance fairness. Consequently, the following objective is constructed:
	\begin{equation} \label{eq:fair_optimization}
		\mathop{\min}\limits_{\boldsymbol \theta} \mathop{\max}\limits_{\boldsymbol \lambda \in \Delta^{K-1}} \sum_{k=1}^{K} \frac{\lambda_k}{N_k} \sum_{(\boldsymbol x,y) \in D_k} \ell(f(\boldsymbol x; \boldsymbol \theta), y),
	\end{equation}
	which requires the model to be fairly optimized on data from different clients. This approach is thus referred to as fair optimization. Several methods have been applied to achieve this, including AFL \cite{mohri2019agnostic}, which directs attention to clients with the largest training loss, and q-FedAvg \cite{li2019fair}, FairFed \cite{ezzeldin2023fairfed}, and FedGA \cite{zhang2023federated}, which assign higher importance to clients with larger training losses. FedCE \cite{jiang2023fair} takes an implicit method by focusing on clients with worse task-specific performance metrics.
	
	Despite some effectiveness, it remains challenging for these methods to fully achieve the objective in Eq. \ref{eq:client_fairness}. As conceptually illustrated in Fig. \ref{fig:motivation}, fair optimization methods may cause the model to quickly converge to sharp minima, as the most significant loss reduction occurs near these points, which Eq. \ref{eq:fair_optimization} favors. However, this rapid convergence does not necessarily guarantee the achievement of Eq. \ref{eq:client_fairness}. The reason for this is the gap between training loss and expected loss, especially for smaller-scale data (\textit{e.g.}, corrupted images) \cite{vapnik1998statistical}. Therefore, it is crucial to broaden the investigation of performance fairness to include generalization, rather than focusing solely on optimization.
	
	\subsubsection{Fair Generalization} \label{sec:fair_generalization}
	Considering this limitation of fair optimization, our preliminary work proposes shifting the focus to another metric \cite{FedISM}. This starts by analyzing the shortcomings of relying solely on training loss. A major issue is its lack of sensitivity to the underlying geometry of the convergence process, treating sharp and flat minima in the same way. This happens because training loss only reflects the 0th-order state of convergence on the loss surface. To address this, this work proposes focusing on the 1st-order state of convergence, which also considers the training loss in its vicinity. The chosen metric is the sharpness of the loss surface \cite{foret2021sharpness, zhuang2022surrogate}, defined as the greatest change in loss when perturbing the initial model parameters:
	\begin{equation} \label{eq:sharpness}
		\mathcal{S} := \max_{\Vert \boldsymbol \epsilon \Vert_2 \leq \rho}{\ell(f(\boldsymbol x; \boldsymbol \theta + \boldsymbol \epsilon), y) -  \ell(f(\boldsymbol x; \boldsymbol \theta), y)},
	\end{equation}
	where $\rho$ is a predefined positive parameter that controls the perturbation’s search distance. Building on this, the modified objective is formulated as:
	\begin{equation} \label{eq:fair_generalization}
		\mathop{\min}\limits_{\boldsymbol \theta} \mathop{\max}\limits_{\boldsymbol \lambda \in \Delta^{K-1}} \sum_{k=1}^{K} \frac{\lambda_k}{N_k} \sum_{(\boldsymbol x,y) \in D_k} \mathcal{S}(f(\boldsymbol x; \boldsymbol \theta), y).
	\end{equation}
	This problem is addressed by the previous method FedISM \cite{FedISM}. In contrast to fair optimization methods that focus on lower bounds of losses (Eq. \ref{eq:fair_optimization}), this method aims to achieve Max-Min fairness in terms of sharpness, promoting convergence to flat minima across all clients’ data, as illustrated in Fig. \ref{fig:motivation}. Since reduced sharpness often correlates with improved generalization \cite{foret2021sharpness, zhuang2022surrogate}, this method places greater emphasis on fairness in terms of generalization rather than optimization alone.
	
	\textbf{\textit{Summary: }} The two categories of methods discussed share a common goal: achieving fairness for a single state of convergence on the loss surface (\textit{i.e.}, aligning the state across clients), as illustrated in Fig. \ref{fig:motivation}. The underlying motivation is to treat the fairness of the single metric as a surrogate for the final fairness objective. Specifically, the first category focuses on the 0th-order state of the convergence, while the second emphasizes the 1st-order state by exploring the region within a fixed search distance. 
	
	\subsection{New Assessment with Comprehensive States} \label{sec:assessment}
	Although the aforementioned two approaches show some effectiveness in addressing fairness problem, we argue that both are suboptimal. This is because a single state does not fully capture the convergence characteristics, making fairness based on this metric (Eq. \ref{eq:fair_optimization} and Eq. \ref{eq:fair_generalization}) an imperfect surrogate for the final fairness objective (Eq. \ref{eq:client_fairness}). In fair optimization, a convergence with uniformly low loss can exhibit varying levels of sharpness, which can be suboptimal for generalization on the test set \cite{FedISM}. For fair generalization, a convergence with uniformly low sharpness (at a fixed search distance) across different distributions can still exhibit high loss and sharpness when considered at other search distances. This leads to suboptimal fairness because the optimal sharpness, which should be minimized for better performance, is defined at different search distances across distributions \cite{DBLP:conf/cvpr/HuangZX00000L23}. Considering their limitations, we ask: how can we better assess a convergence than by using a single state (\textit{i.e.}, loss and sharpness), so that we can leverage this assessment to guide fair learning?
	
	To address this issue, we extend the focus to multiple states. Specifically, we use a series of sharpness (or its variants) defined at different search distances $\rho \in [0, \rho_{max}]$, where $\rho_{max}$ is a predefined maximum.
	
	We begin by defining sharpness as a function of $\rho$:
	\begin{equation} \label{eq:sharpness_f}
		\mathcal{S}(\rho) := \max_{\Vert \boldsymbol \epsilon \Vert_2 \leq \rho}\{\ell(f(\boldsymbol x; \boldsymbol \theta + \boldsymbol \epsilon), y) -  \ell(f(\boldsymbol x; \boldsymbol \theta), y)\}.
	\end{equation}
	This term is difficult to compute directly due to the continuous and unbounded nature of the perturbation. To make this more manageable, we approximate the difference linearly using the Taylor series expansion, assuming $\rho$ is always small enough:
	\begin{equation} \label{eq:approximation}
		\ell(f(\boldsymbol x; \boldsymbol \theta + \boldsymbol \epsilon), y) -  \ell(f(\boldsymbol x; \boldsymbol \theta), y) \approx \boldsymbol \epsilon^\top \nabla \ell(f(\boldsymbol x; \boldsymbol \theta), y).
	\end{equation}
	Using this approximation, we can determine the optimal perturbation by maximizing the right-hand side expression:
	\begin{equation} \label{eq:optimal_perturbation}
		\boldsymbol \epsilon^* = \mathop{\arg\max}\limits_{\Vert \boldsymbol \epsilon \Vert_2 \leq \rho} \boldsymbol \epsilon^\top \nabla \ell(f(\boldsymbol x; \boldsymbol \theta), y) = \rho \frac{\nabla \ell(f(\boldsymbol x; \boldsymbol \theta), y)}{\Vert\nabla \ell(f(\boldsymbol x; \boldsymbol \theta), y) \Vert_2}.
	\end{equation}
	It is clear that $\boldsymbol \epsilon^*$ is a function of $\rho$, and thus it is denoted as $\boldsymbol \epsilon^*(\rho)$. In this way, sharpness can be calculated more feasibly.
	\begin{equation} \label{eq:sharpness_approximation}
		\mathcal{S}(\rho) \approx  \ell(f(\boldsymbol x; \boldsymbol \theta+ \boldsymbol \epsilon^*(\rho)), y) -  \ell(f(\boldsymbol x; \boldsymbol \theta), y).
	\end{equation}
	In addition to sharpness, we explore its variant, the perturbed loss, which also reflects multiple states:
	\begin{equation} \label{eq:loss_p}
		\mathcal{L}(\rho) \approx  \ell(f(\boldsymbol x; \boldsymbol \theta+ \boldsymbol \epsilon^*(\rho)), y).
	\end{equation}
	
	Please note that these two quantities, \textit{i.e.}, $\mathcal{S}$ and $\mathcal{L}$, vary as $\rho$ changes. Therefore, we can assess the convergence by a series of sharpness or perturbed loss, \textit{i.e.}, $\{\mathcal{S}(\rho)  \mid \rho \in [0, \rho_{max}]\}$ or $\{\mathcal{L}(\rho)  \mid \rho \in [0, \rho_{max}]\}$.
	This assessment is, in fact, a generalized version of those using the single state discussed in Sec. \ref{sec:previous_solutions}, as it not only includes the 0th-order state of convergence when $\rho = 0$, but also considers the 1st-order state defined at varying distances $\rho > 0$.

	\begin{figure}[!t] 
		\centering
		\includegraphics[width=1.0\columnwidth]{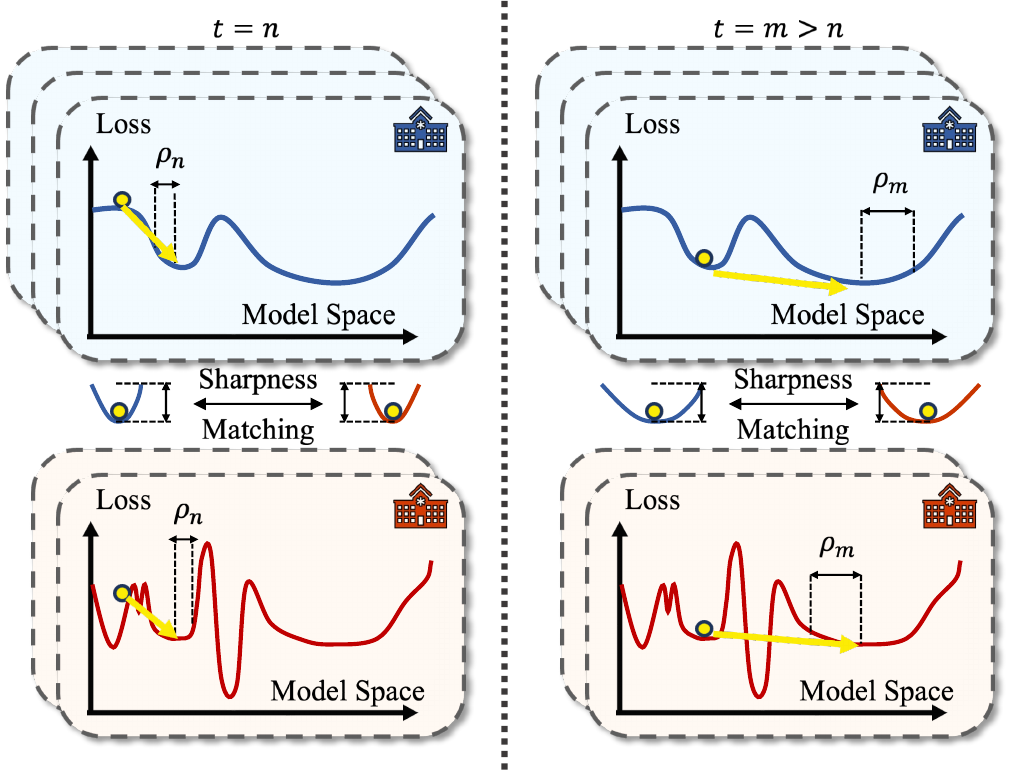}
		\caption{Illustration of FedISM+. The key insight is to maintain uniformly low sharpness across clients, defined at progressively increasing search distances as training progresses.}
		\label{fig:framework}
	\end{figure}

	\subsection{FedISM+: Extending State Matching Further} \label{sec:FedISM+}
	
	Building on the comprehensive assessment discussed above, we extend the previous single-state matching methods from Sec. \ref{sec:previous_solutions}. Our insight is to achieve Max-Min fairness for these comprehensive states across clients, rather than relying on the single state as in Eq. \ref{eq:fair_optimization} and Eq. \ref{eq:fair_generalization}, which better promotes fairness on testing sets.
	
	To achieve this, we propose an enhanced method, called inter-client progressive state matching (\textbf{FedISM+}). The overview is depicted in Fig. \ref{fig:framework}. Building on the previous method, FedISM \cite{FedISM}, the key feature of FedISM+ is its progressive scheme that focuses on fairness across multiple states of convergence. This scheme is implemented by defining a search distance for computing sharpness (Eq. \ref{eq:sharpness_approximation}) and perturbed loss (Eq. \ref{eq:loss_p}), as follows:
	\begin{equation} \label{eq:rho_scheme}
		\rho(t) = \rho_{max} \left( \frac{t}{T} \right)^\tau,
	\end{equation}
	where $\rho_{max}$ and $\tau > 0$ are predefined parameters, $t$ denotes the current round, and $T$ represents the maximum communication round in FL. This design ensures that the search distance increases from 0 to $\rho_{max}$ during training, controlled by $\tau$. Initially, this scheme effectively considers the 0th-order state since $\rho(t) \approx 0$, and progressively incorporates the 1st-order state as training progresses.
	
	\setlength{\fboxrule}{1pt}
	\begin{algorithm}[!t]
		\caption{\colorbox[rgb]{1.0, 0.90, 0.87}{$\texttt{FedAvg}$} and \colorbox[rgb]{0.87, 0.90, 1}{$\texttt{FedISM+}$}}
		\label{alg:FedISM+}
		\textbf{Input}: Number of clients $K$, local datasets $\{D_1, \dots, D_K\}$, local dataset size, total communication rounds $T$, learning rate of local training $\eta$, maximum search distance $\rho_{max}$, parameter $\tau$.\\
		\textbf{Output}: Global model $\boldsymbol \theta_{T+1}$
		\begin{algorithmic}[1] %[1] enables line numbers
			\STATE Initialize the global model $\boldsymbol \theta_1$
			\FOR{$t = 1, 2, \dots, T$}
			\STATE \colorbox[rgb]{0.87, 0.90, 1}{$\rho \leftarrow \rho_{max} \left( \frac{t}{T} \right)^\tau$} \hfill \textcolor{blue}{$\triangleright$ set $\rho$ for $\texttt{FedISM+}$}
			\FOR{Client $k = 1, 2, \dots, K$ in parallel}
			\STATE $\boldsymbol \theta_{(t,k)} \leftarrow \boldsymbol \theta_t$ \hfill \textcolor{blue}{$\triangleright$ download the global model}
			\FOR{$(\boldsymbol x_i, y_i) \in D_k$}
			\STATE \colorbox[rgb]{1.0, 0.90, 0.87}{Update $\boldsymbol \theta_{(t,k)}$ with $(\boldsymbol x_i, y_i)$ by Eq. \ref{eq:GD}}
			\STATE \colorbox[rgb]{0.87, 0.90, 1}{Update $\boldsymbol \theta_{(t,k)}$ with $(\boldsymbol x_i, y_i)$ by Eq. \ref{eq:SAM}}
			\ENDFOR
			\ENDFOR
			\STATE \colorbox[rgb]{1.0, 0.90, 0.87}{$\boldsymbol \theta_{t+1} \leftarrow$ Aggregate $\{\boldsymbol \theta_{(t,k)}\}_{k=1}^K$ with $\boldsymbol w_{\texttt{Avg}}$ (Eq. \ref{eq:FedAvg_weights})}
			\STATE \colorbox[rgb]{0.87, 0.90, 1}{$\boldsymbol \theta_{t+1} \leftarrow$ Aggregate $\{\boldsymbol \theta_{(t,k)}\}_{k=1}^K$ with $\boldsymbol w_t$ (Eq. \ref{eq:FedISM_weights_EMA})}
			\ENDFOR
			\RETURN $\boldsymbol \theta_{T+1}$
		\end{algorithmic}
	\end{algorithm}
	
	\begin{table*}[!t]
		\centering
		\renewcommand{\arraystretch}{1.0}
		\caption{Component-wise ablation study. Numbers outside the parentheses correspond to the mean (\%), while those within the parentheses are the standard deviation (\%). $\Delta$ACC (\%) and $\Delta$AUC (\%) denote the difference to ACC and AUC of FedAvg. The best results are in bold.}
		\resizebox{1.0\textwidth}{!}{
			\begin{tabular}{c|c|ccc|cccccccccccc}
				\toprule
				\hline
				\multirow{3}{*}{Type} & \multirow{3}{*}{\begin{tabular}[c]{@{}c@{}}Method\\ Name\end{tabular}} & \multicolumn{3}{c|}{\multirow{2}{*}{Component}}                                               & \multicolumn{12}{c}{RSNA ICH}                                                                                                                                                                                                                                                                                             \\ \cline{6-17} 
				&                                                                        & \multicolumn{3}{c|}{}                                                                         & \multicolumn{4}{c|}{Clean}                                                                                           & \multicolumn{4}{c|}{Crorrupted}                                                                                       & \multicolumn{4}{c}{Average}                                                                     \\ \cline{3-17} 
				&                                                                        & SALT                          & SAGA (S)                      & SAGA (L)                      & ACC $\uparrow$ & $\Delta$ACC $\uparrow$        & AUC $\uparrow$ & \multicolumn{1}{c|}{$\Delta$AUC $\uparrow$}        & ACC $\uparrow$ & $\Delta$ACC $\uparrow$         & AUC $\uparrow$ & \multicolumn{1}{c|}{$\Delta$AUC $\uparrow$}        & ACC $\uparrow$ & $\Delta$ACC $\uparrow$        & AUC $\uparrow$ & $\Delta$AUC $\uparrow$        \\ \hline \hline
				\multirow{2}{*}{-}           & \multirow{2}{*}{FedAvg \cite{mcmahan2017communication}}                                                & \multirow{2}{*}{}             & \multirow{2}{*}{}             & \multirow{2}{*}{}             & $76.77$        & \multirow{2}{*}{-}            & $94.58$        & \multicolumn{1}{c|}{\multirow{2}{*}{-}}            & $53.66$        & \multirow{2}{*}{-}             & $84.37$        & \multicolumn{1}{c|}{\multirow{2}{*}{-}}            & $65.21$        & \multirow{2}{*}{-}            & $89.48$        & \multirow{2}{*}{-}            \\ 
				&                                                                        &                               &                               &                               & $(0.68)$       &                               & $(0.20)$       & \multicolumn{1}{c|}{}                              & $(1.28)$       &                                & $(0.73)$       & \multicolumn{1}{c|}{}                              & $(0.75)$       &                               & $(0.34)$       &                               \\ \hline \hline
				\multirow{10}{*}{ {\begin{tabular}[c]{@{}c@{}}Ablation\\ of \\FedISM+\end{tabular}} }    & \multirow{2}{*}{-}                                                     & \multirow{2}{*}{$\checkmark$} & \multirow{2}{*}{}             & \multirow{2}{*}{}             & $\bm{78.84}$        & \multirow{2}{*}{$\bm{+2.07}$}      & $\bm{95.64}$        & \multicolumn{1}{c|}{\multirow{2}{*}{$\bm{+1.06}$}}      & $53.53$        & \multirow{2}{*}{$-0.13$}       & $86.50$        & \multicolumn{1}{c|}{\multirow{2}{*}{$+2.13$}}      & $66.19$        & \multirow{2}{*}{$+0.98$}      & $91.07$        & \multirow{2}{*}{$+1.59$}      \\
				&                                                                        &                               &                               &                               & $\bm{(0.72)}$       &                               & $\bm{(0.08)}$       & \multicolumn{1}{c|}{}                              & $(2.79)$       &                                & $(0.94)$       & \multicolumn{1}{c|}{}                              & $(1.11)$       &                               & $(0.44)$       &                               \\ \cline{2-17} 
				& \multirow{2}{*}{-}                                                     & \multirow{2}{*}{}             & \multirow{2}{*}{$\checkmark$} & \multirow{2}{*}{}             & $74.48$        & \multirow{2}{*}{$-2.29$}      & $93.73$        & \multicolumn{1}{c|}{\multirow{2}{*}{$-0.85$}}      & $59.16$        & \multirow{2}{*}{$+5.50$}       & $86.96$        & \multicolumn{1}{c|}{\multirow{2}{*}{$+2.59$}}      & $66.82$        & \multirow{2}{*}{$+1.61$}      & $90.34$        & \multirow{2}{*}{$+0.86$}      \\
				&                                                                        &                               &                               &                               & $(1.73)$       &                               & $(0.54)$       & \multicolumn{1}{c|}{}                              & $(2.42)$       &                                & $(0.94)$       & \multicolumn{1}{c|}{}                              & $(1.06)$       &                               & $(0.34)$       &                               \\ \cline{2-17} 
				& \multirow{2}{*}{-}                                                     & \multirow{2}{*}{}             & \multirow{2}{*}{}             & \multirow{2}{*}{$\checkmark$} & $73.44$        & \multirow{2}{*}{$-3.33$}      & $93.38$        & \multicolumn{1}{c|}{\multirow{2}{*}{$-1.20$}}      & $61.39$        & \multirow{2}{*}{$+7.73$}       & $87.68$        & \multicolumn{1}{c|}{\multirow{2}{*}{$+3.31$}}      & $67.42$        & \multirow{2}{*}{$+2.21$}      & $90.53$        & \multirow{2}{*}{$+1.05$}      \\
				&                                                                        &                               &                               &                               & $(1.59)$       &                               & $(0.43)$       & \multicolumn{1}{c|}{}                              & $(1.50)$       &                                & $(0.36)$       & \multicolumn{1}{c|}{}                              & $(0.81)$       &                               & $(0.22)$       &                               \\ \cline{2-17} 
				& \multirow{2}{*}{FedISM+ (S)}                                           & \multirow{2}{*}{$\checkmark$} & \multirow{2}{*}{$\checkmark$} & \multirow{2}{*}{}             & $77.19$        & \multirow{2}{*}{$+0.42$}      & $95.01$        & \multicolumn{1}{c|}{\multirow{2}{*}{$+0.43$}}      & $\bm{64.55}$   & \multirow{2}{*}{$\bm{+10.89}$} & $\bm{89.93}$   & \multicolumn{1}{c|}{\multirow{2}{*}{$\bm{+5.56}$}} & $70.87$        & \multirow{2}{*}{$+5.66$}      & $92.47$        & \multirow{2}{*}{$+2.99$}      \\
				&                                                                        &                               &                               &                               & $(0.67)$       &                               & $(0.10)$       & \multicolumn{1}{c|}{}                              & $\bm{(0.78)}$  &                                & $\bm{(0.22)}$  & \multicolumn{1}{c|}{}                              & $(0.42)$       &                               & $(0.09)$       &                               \\ \cline{2-17} 
				& \multirow{2}{*}{FedISM+ (L)}                                           & \multirow{2}{*}{$\checkmark$} & \multirow{2}{*}{}             & \multirow{2}{*}{$\checkmark$} & $78.09$        & \multirow{2}{*}{$+1.32$}      & $95.31$        & \multicolumn{1}{c|}{\multirow{2}{*}{$+0.73$}}      & $63.83$        & \multirow{2}{*}{$+10.17$}      & $89.72$        & \multicolumn{1}{c|}{\multirow{2}{*}{$+5.35$}}      & $\bm{70.96}$        & \multirow{2}{*}{$\bm{+5.75}$}      & $\bm{92.51}$   & \multirow{2}{*}{$\bm{+3.03}$} \\
				&                                                                        &                               &                               &                               & $(0.78)$       &                               & $(0.14)$       & \multicolumn{1}{c|}{}                              & $(1.02)$       &                                & $(0.28)$       & \multicolumn{1}{c|}{}                              & $\bm{(0.70)}$       &                               & $\bm{(0.17)}$  &                               \\ \hline
				\bottomrule
			\end{tabular}
		}
		\label{tab:Ablation}
	\end{table*}
	
	Next, we describe how fairness is achieved for these states. In traditional local training within FL (\textit{e.g.}, FedAvg \cite{mcmahan2017communication}), the objective is to minimize the local loss via gradient descent:
	\begin{equation} \label{eq:GD}
		\boldsymbol \theta \leftarrow \boldsymbol \theta - \eta \nabla \ell(f(\boldsymbol x; \boldsymbol \theta), y),
	\end{equation}
	where $\eta$ is the learning rate. However, this method only considers the 0th-order state and does not account for other states, \textit{i.e.} $\mathcal{S}(\rho(t))$ and $\mathcal{L}(\rho(t))$. To overcome this limitation, we introduce sharpness-awareness into local training. Drawing from sharpness-aware minimization \cite{foret2021sharpness}, we modify the update rule to:
	\begin{equation} \label{eq:SAM}
		\boldsymbol \theta \leftarrow \boldsymbol \theta - \eta \nabla \ell(f(\boldsymbol x; \boldsymbol \theta + \boldsymbol \epsilon^*(\rho(t))), y),
	\end{equation}
	where $\boldsymbol \epsilon^*(\rho(t))$ refers to the optimal perturbation that maximizes the change in training loss, as outlined in Eq. \ref{eq:optimal_perturbation}. This adjustment guides the optimization process to a destination where the neighboring region has a small loss. The perturbation is a function of $\rho(t)$, which allows for progressively exploring the neighboring region in a comprehensive manner. For simplicity, we refer to this as sharpness-aware local training (SALT).
	
	After each local training round, the gradients from all participating clients are sent to the server for aggregation. Achieving fairness in the states of convergence across clients depends on whether these gradients are aggregated fairly. In vanilla FL \cite{mcmahan2017communication}, aggregation weights are determined by the size of the dataset each client holds, as expressed by:
	\begin{equation} \label{eq:FedAvg_weights}
		\boldsymbol w_{\texttt{Avg}} = \frac{1}{ \sum_{k=1}^{K} N_k}[N_1, N_2, \cdots, N_K]^\top.
	\end{equation}
	However, this weighting scheme may not achieve our fairness goal. As discussed in Fig. \ref{fig:background}, clients with corrupted data typically have fewer data points, leading to the global model update being predominantly driven by clients with clean data. As a result, sharpness and perturbed loss are minimized mainly for clients with clean data, causing a disparity in these metrics that mirrors the loss disparity in fair optimization \cite{mohri2019agnostic, li2019fair, ezzeldin2023fairfed, jiang2023fair}. To address this issue, we introduce sharpness-aware global aggregation (SAGA), where the aggregation weights are determined by sharpness, as follows:
	\begin{equation} \label{eq:FedISM_weights}
		\widetilde{\boldsymbol w}_{t} = \frac{1}{ \sum_{k=1}^{K} \mathcal{S}_{k,t}^{(q)}}[\mathcal{S}_{1,t}^{(q)}, \mathcal{S}_{2,t}^{(q)}, \cdots, \mathcal{S}_{K,t}^{(q)}]^\top,
	\end{equation}
	Alternatively, we can use the perturbed loss as the weighting criterion:
	\begin{equation} \label{eq:FedISM+_weights}
		\widetilde{\boldsymbol w}_{t} = \frac{1}{ \sum_{k=1}^{K} \mathcal{L}_{k,t}^{(q)}} [\mathcal{L}_{1,t}^{(q)}, \mathcal{L}_{2,t}^{(q)}, \cdots, \mathcal{L}_{K,t}^{(q)}]^\top,
	\end{equation}
	where $\mathcal{S}_{k,t}$ and $\mathcal{L}_{k,t}$ represent the sharpness and perturbed loss computed for the complete local dataset $D_k$ at round $t$ using Eq. \ref{eq:sharpness_approximation} and Eq. \ref{eq:loss_p}, respectively, with $q > 0$ as the exponent. Methods using these two aggregation weights (denoted as SAGA (S) and SAGA (L)) are referred to as FedISM+ (S) and FedISM+ (L), respectively. This strategy assigns greater weight to clients with higher sharpness or perturbed loss during aggregation. Importantly, when using SALT (Eq. \ref{eq:SAM}), each client’s gradient inherently points towards reducing the sharpness specific to its local data. Therefore, the combination of SALT and SAGA prioritizes minimizing the sharpness/perturbed loss for clients with initially higher sharpness/perturbed loss, effectively promoting fairness in these progressive states. To ensure the stability of FL, a simple moving average is further applied for rounds $t > 1$, as follows:
	\begin{equation} \label{eq:FedISM_weights_EMA}
		{\boldsymbol w}_{t} = \beta \widetilde{\boldsymbol w}_{t} + (1-\beta) {\boldsymbol w}_{t-1}.
	\end{equation}
	Particularly, we set ${\boldsymbol w}_{1} = \widetilde{\boldsymbol w}_{1}$ for the first round.
	
	It is important to note that FedISM+ only requires clients to transmit their sharpness or perturbed loss, not any information about their data distributions. This preserves privacy, as discussed in Sec. \ref{sec:privacy}.
	
	For clarity, the procedure for FedISM+ is summarized in Alg. \ref{alg:FedISM+}, where we also include FedAvg \cite{mcmahan2017communication} for better understanding. It shows that FedISM+ introduces modifications only to the local optimizer and the global aggregation weights compared to FedAvg, along with a scheme to determine the search distance, avoiding the introduction of complex loss functions or regularization methods. This pseudocode illustrates the ease of implementation.

	\begin{table*}[!t]
		\centering
		\caption{Performance comparison on RSNA ICH. Numbers outside the parentheses correspond to the mean (\%), while those within the parentheses are the standard deviation (\%). The best results are in bold.}
		\resizebox{1.0\textwidth}{!}{
			\renewcommand{\arraystretch}{1.0}
			\begin{tabular}{c|c|cccccccccccc}
				\toprule
				\hline
				\multirow{4}{*}{Category} & \multirow{4}{*}{Method}      & \multicolumn{12}{c}{RSNA ICH}                                                                                                                                                                       \\ \cline{3-14} 
				&                              & \multicolumn{6}{c|}{Gaussian Noise}                                                                  & \multicolumn{6}{c}{Motion Blur}                                                         \\ \cline{3-14} 
				&                              & \multicolumn{2}{c}{Clean} & \multicolumn{2}{c}{Corrupted} & \multicolumn{2}{c|}{Average}             & \multicolumn{2}{c}{Clean} & \multicolumn{2}{c}{Corrupted} & \multicolumn{2}{c}{Average} \\ \cline{3-14} 
				&                              & ACC $\uparrow$         & AUC $\uparrow$         & ACC $\uparrow$           & AUC $\uparrow$           & ACC $\uparrow$      & \multicolumn{1}{c|}{AUC $\uparrow$}      & ACC $\uparrow$         & AUC $\uparrow$         & ACC $\uparrow$           & AUC $\uparrow$           & ACC $\uparrow$          & AUC $\uparrow$          \\ \hline \hline
				\multirow{2}{*}{Vanilla FL} & FedAvg \cite{mcmahan2017communication}                       & $76.77$     & $94.58$     & $53.66$       & $84.37$       & $65.21$  & \multicolumn{1}{c|}{$89.48$}  & $75.97$     & $94.74$     & $68.69$       & $90.98$       & $72.33$      & $92.86$      \\
				& (AISTATS'17)                 & $(0.68)$    & $(0.20)$    & $(1.28)$      & $(0.73)$      & $(0.75)$ & \multicolumn{1}{c|}{$(0.34)$} & $(1.11)$    & $(0.22)$    & $(0.74)$      & $(0.41)$      & $(0.81)$     & $(0.29)$     \\ \hline
				\hline
				\multirow{10}{*}{\begin{tabular}[c]{@{}c@{}}Fair Optimization\\ (0th-Order State)\end{tabular}} & Agnostic-FL  \cite{mohri2019agnostic}                  & $55.13$     & $83.20$      & $46.69$       & $78.13$       & $50.91$  & \multicolumn{1}{c|}{$80.67$}  & $63.57$     & $89.45$     & $57.27$       & $86.35$       & $60.42$      & $87.90$       \\
				& (ICML'19)                    & $(7.28)$    & $(4.86)$    & $(7.54)$      & $(4.79)$      & $(2.56)$ & \multicolumn{1}{c|}{$(1.65)$} & $(3.48)$    & $(1.32)$    & $(7.12)$      & $(3.32)$      & $(3.27)$     & $(1.78)$     \\ \cline{2-14} 
				& q-FedAvg \cite{li2019fair}                     & $75.94$     & $94.36$     & $59.46$       & $86.81$       & $67.70$   & \multicolumn{1}{c|}{$90.58$}  & $75.68$     & $94.77$     & $70.53$       & $91.87$       & $73.10$       & $93.32$      \\
				& (ICLR'20)                    & $(1.05)$    & $(0.28)$    & $(1.90)$      & $(0.95)$      & $(0.69)$ & \multicolumn{1}{c|}{$(0.40)$} & $(1.59)$    & $(0.34)$    & $(1.18)$      & $(0.53)$      & $(0.54)$     & $(0.17)$     \\ \cline{2-14} 
				& FairFed \cite{ezzeldin2023fairfed}                      & $74.13$     & $93.55$     & $60.27$       & $86.74$       & $67.20$  & \multicolumn{1}{c|}{$90.15$}  & $74.59$     & $94.34$     & $69.66$       & $91.75$       & $72.12$      & $93.05$      \\
				& (AAAI'23)                    & $(1.15)$    & $(0.30)$    & $(1.08)$      & $(0.63)$      & $(0.79)$ & \multicolumn{1}{c|}{$(0.38)$} & $(1.12)$    & $(0.21)$    & $(1.08)$      & $(0.30)$      & $(0.66)$     & $(0.18)$     \\ \cline{2-14} 
				& FedCE \cite{jiang2023fair}                        & $75.82$     & $94.31$     & $58.77$       & $86.93$       & $67.29$  & \multicolumn{1}{c|}{$90.62$}  & $76.83$     & $95.00$     & $70.38$       & $92.04$       & $73.60$      & $93.52$      \\
				& (CVPR'23)                    & $(0.45)$    & $(0.10)$    & $(1.92)$      & $(0.70)$      & $(0.94)$ & \multicolumn{1}{c|}{$(0.37)$} & $(0.83)$    & $(0.15)$    & $(0.72)$      & $(0.28)$      & $(0.73)$     & $(0.17)$     \\ \cline{2-14} 
				& FedGA \cite{zhang2023federated}                        & $71.93$     & $92.88$     & $60.73$       & $87.45$       & $66.33$  & \multicolumn{1}{c|}{$90.16$}  & $72.97$     & $93.75$     & $69.64$       & $91.78$       & $71.31$      & $92.76$      \\
				& (CVPR'23)                    & $(1.43)$    & $(0.36)$    & $(1.24)$      & $(0.47)$      & $(0.99)$ & \multicolumn{1}{c|}{$(0.33)$} & $(0.78)$    & $(0.16)$    & $(0.60)$      & $(0.27)$      & $(0.54)$     & $(0.16)$     \\ \hline \hline
				
				\multirow{2}{*}{\begin{tabular}[c]{@{}c@{}}Fair Generalization\\\ (1st-Order State)\end{tabular}}     & FedISM \cite{FedISM}                   & $77.52$     & $95.02$     & $64.51$       & $89.56$       & $\bm{71.01}$  & \multicolumn{1}{c|}{$92.29$}  & $75.43$     & $95.06$     & $72.54$       & $93.46$       & $73.99$      & $94.26$      \\
				& (IJCAI'24)                   & $(0.58)$    & $(0.16)$    & $(0.64)$      & $(0.22)$      & $\bm{(0.36)}$ & \multicolumn{1}{c|}{$(0.13)$} & $(0.97)$    & $(0.11)$    & $(0.79)$      & $(0.15)$      & $(0.54)$     & $(0.12)$     \\ \hline \hline
				\multirow{4}{*}{\begin{tabular}[c]{@{}c@{}}Ours\\ (Multiple States)\end{tabular}}                          & \multirow{2}{*}{FedISM+ (S)} & $77.19$     & $95.01$     & $\bm{64.55}$       & $\bm{89.93}$       & $70.87$  & \multicolumn{1}{c|}{$92.47$}  & $76.50$     & $95.17$     & $\bm{72.97}$       & $93.75$       & $74.73$      & $94.46$      \\
				&                              & $(0.67)$    & $(0.10)$    & $\bm{(0.78)}$      & $\bm{(0.22)}$      & $(0.42)$ & \multicolumn{1}{c|}{$(0.09)$} & $(0.70)$    & $(0.07)$    & $\bm{(0.60)}$      & $(0.10)$      & $(0.34)$     & $(0.05)$     \\ \cline{2-14} 
				& \multirow{2}{*}{FedISM+ (L)} & $\bm{78.09}$     & $\bm{95.31}$     & $63.83$       & $89.72$       & $70.96$  & \multicolumn{1}{c|}{$\bm{92.51}$}  & $\bm{77.60}$     & $\bm{95.50}$      & $72.88$       & $\bm{93.89}$       & $\bm{75.24}$      & $\bm{94.70}$      \\
				&                              & $\bm{(0.78)}$    & $\bm{(0.14)}$    & $(1.02)$      & $(0.28)$      & $(0.70)$ & \multicolumn{1}{c|}{$\bm{(0.17)}$} & $\bm{(0.58)}$    & $\bm{(0.07)}$    & $(0.64)$      & $\bm{(0.11)}$      & $\bm{(0.42)}$     & $\bm{(0.06)}$     \\ \hline
				\bottomrule
			\end{tabular}
		}
		\label{tab:SOTA-ICH}
	\end{table*}

	\section{Experiments} \label{sec:eval}
	We conduct a range of experiments to assess the effectiveness of our proposed solution.
	
	\subsection{Experimental Setup}
	\subsubsection{Datasets}
	Two widely-used medical image classification datasets, commonly employed in FL research \cite{jiang2022dynamic, DBLP:conf/ijcai/Wu0JCY23}, are used for evaluation:
	\begin{itemize}
		\item RSNA ICH \cite{flanders2020construction}: The task involves classifying CT slices into five intracranial hemorrhage subtypes. Following \cite{jiang2022dynamic}, 25,000 images are randomly selected for experiments.
		\item ISIC 2019 \cite{tschandl2018ham10000,codella2018skin,combalia2019bcn20000}: This dataset contains 25331 images for training models to classify eight skin diseases.
	\end{itemize}
	Both datasets are divided into training and test sets in an 8:2 ratio and resized to $224 \times 224$ pixels, in line with the standard preprocessing steps \cite{jiang2022dynamic}.
	
	\subsubsection{Client Training Data Preparation}
	Following \cite{xu2022fedcorr}, the training sets are distributed among 20 clients using a Dirichlet distribution (\textit{i.e.}, Dir(1.0)), simulating prevalent label distribution shifts. To create imaging quality shifts, Gaussian noise or motion blur is added to images for a subset of the clients, as depicted in Fig. \ref{fig:background}, following \cite{hendrycks2019benchmarking}.
	
	\subsubsection{Model}
	Pretrained ResNet-18 \cite{he2016deep} is selected as the base model for all experiments for standard evaluation.
	
	\subsubsection{Implement Details}
	To mitigate label distribution shifts, we incorporate logit adjustment \cite{menon2020long} in local training. Following previous settings \cite{FedISM}, we train the local models using batches of 32 images with the Adam optimizer, applying a constant learning rate of 0.0003, beta values of (0.9, 0.999), and a weight decay of 0.0005. For FL setup, we configure a maximum of 300 communication rounds and set the local epoch to 1. These hyperparameters are kept consistent across all experiments to ensure a fair comparison. For FedISM+, we adopt GSAM \cite{zhuang2022surrogate} for sharpness-aware minimization and set $\rho_{max}$ = 0.1, $\tau$ = 0.5, $q$ = 2.0 and $\beta$ = 0.5 defaultly.

	\subsubsection{Evaluation Strategy}
	We follow the previous work \cite{FedISM} to design the evaluation strategy.
	To maximize the utilization of limited testing images, the testing set is not directly divided among clients. Instead, following \cite{hendrycks2019benchmarking}, a new corrupted testing set is generated from the initial clean one by adding the same type of corruption that is applied to the training set of corrupted clients. The performance of FL is then evaluated on both the original clean and the newly generated corrupted test sets to demonstrate how Eq. \ref{eq:objective} is achieved. We use the area under the receiver operating characteristic curve (AUC) and classification accuracy (ACC) as our evaluation metrics. To minimize the impact of randomness, all experiments are independently conducted three times. We present the mean and standard deviation calculated from the last five communication rounds, following the strategy in \cite{huang2023rethinking}.

	\subsection{Ablation Study} \label{sec:component_ablation}
	FedISM+ consists of two primary components: SALT and SAGA. To assess the impact of each component, we perform ablation studies on the RSNA ICH dataset, using 20\% of the clients with Gaussian noise corruption, and integrate each component individually with FedAvg \cite{mcmahan2017communication}. The results of AUC and ACC for both clean and corrupted images, as well as their average, are summarized in Tab. \ref{tab:Ablation}. SALT, which encourages the model to converge towards flatter minima, typically improves the performance of FedAvg. However, it does not fully lead to fairness, as it overlooks the cross-client difference. This problem is addressed by SAGA. The results demonstrate that both SAGA (S) and SAGA (L) effectively prioritize worst-performing distributions, often those corresponding to corrupted images. Therefore, combining both SALT and SAGA achieves the best results on corrupted images, showcasing the comprehensive benefits of the FedISM+.

	\subsection{Comparison to State-of-the-Arts} \label{sec:sota}
	To demonstrate the superiority, we compare FedISM+ against several state-of-the-art methods, including:
	\begin{itemize}
		\item FedAvg \cite{mcmahan2017communication}: The vanilla FL approach.
		
		\item Agnostic-FL \cite{mohri2019agnostic}: A pioneering work in fair FL that focuses updates on the poorest-performing client.
		
		\item q-FedAvg \cite{li2019fair}: Integrates training loss into the global aggregation stage, with an adjustment including a multiplicative constant for improved convergence. Its parameter $q$ is tuned across the values \{0.5, 1.0, 2.0, 5.0\}.
		
		\item FairFed \cite{ezzeldin2023fairfed}: Seeks to achieve group fairness by optimizing across clients in a balanced manner, with the parameter $\beta$ varied within the range \{0.1, 0.5, 1.0\}.
		
		\item FedCE \cite{jiang2023fair}: Aims to foster fairness in FL, specifically for medical image segmentation. Adapted for classification tasks in this study.
		
		\item FedGA \cite{zhang2023federated}: Focuses on improving domain generalization by ensuring fairness across diverse clients.
		
		\item FedISM \cite{FedISM}: Aims for fairness in sharpness that is defined at a fixed search distance. This is the previous version of this work. We set $q$ = 2.0 and $\rho$ = 0.05 as the origin paper.
	\end{itemize}
	Among these methods, Agnostic-FL \cite{mohri2019agnostic}, q-FedAvg \cite{li2019fair}, FairFed \cite{ezzeldin2023fairfed}, FedCE \cite{jiang2023fair}, and FedGA \cite{zhang2023federated} focus on the fairness of the 0th-order state of convergence, corresponding to fair optimization, while FedISM \cite{FedISM} focuses on the fairness of the 1st-order state of convergence, corresponding to fair generalization.
	
	In this section, experiments are conducted on two datasets in a setting where 4 out of 20 clients (corruption ratio: 20\%) possess images corrupted by Gaussian noise or motion blur.

	\begin{table*}[!t]
		\centering
		\caption{Performance comparison on ISIC 2019. Numbers outside the parentheses correspond to the mean (\%), while those within the parentheses are the standard deviation (\%). The best results are in bold.}
		\resizebox{1.0\textwidth}{!}{
			\renewcommand{\arraystretch}{1.0}
			\begin{tabular}{c|c|cccccccccccc}
				\toprule
				\hline
				\multirow{4}{*}{Category} & \multirow{4}{*}{Method}      & \multicolumn{12}{c}{ISIC 2019}                                                                                                                                                                 \\ \cline{3-14} 
				&                              & \multicolumn{6}{c|}{Gaussian Noise}                                                                  & \multicolumn{6}{c}{Motion Blur}                                                         \\ \cline{3-14} 
				&                              & \multicolumn{2}{c}{Clean} & \multicolumn{2}{c}{Corrupted} & \multicolumn{2}{c|}{Average}             & \multicolumn{2}{c}{Clean} & \multicolumn{2}{c}{Corrupted} & \multicolumn{2}{c}{Average} \\ \cline{3-14} 
				&                              & ACC $\uparrow$         & AUC $\uparrow$         & ACC $\uparrow$           & AUC $\uparrow$           & ACC $\uparrow$      & \multicolumn{1}{c|}{AUC $\uparrow$}      & ACC $\uparrow$         & AUC $\uparrow$         & ACC $\uparrow$           & AUC $\uparrow$           & ACC $\uparrow$          & AUC $\uparrow$          \\ \hline \hline
				\multirow{2}{*}{Vanilla FL} & FedAvg \cite{mcmahan2017communication}                       & $64.43$     & $91.91$     & $38.26$       & $78.03$       & $51.35$  & \multicolumn{1}{c|}{$84.97$}  & $67.16$     & $92.60$     & $54.98$       & $88.16$       & $61.07$      & $90.38$      \\
				& (AISTATS'17)                 & $(1.01)$    & $(0.40)$    & $(1.57)$      & $(1.32)$      & $(0.70)$ & \multicolumn{1}{c|}{$(0.64)$} & $(1.15)$    & $(0.37)$    & $(1.72)$      & $(0.54)$      & $(1.31)$     & $(0.41)$     \\ \hline \hline
				\multirow{10}{*}{\begin{tabular}[c]{@{}c@{}}Fair Optimization\\ (0th-Order State)\end{tabular}} & Agnostic-FL \cite{mohri2019agnostic}                  & $39.66$     & $78.64$     & $33.55$       & $75.62$       & $36.60$  & \multicolumn{1}{c|}{$77.13$}  & $51.24$     & $87.90$     & $50.11$       & $86.56$       & $50.67$      & $87.23$      \\
				& (ICML'19)                    & $(2.94)$    & $(9.08)$    & $(8.88)$      & $(6.48)$      & $(3.30)$ & \multicolumn{1}{c|}{$(2.52)$} & $(5.49)$    & $(1.91)$    & $(3.85)$      & $(1.64)$      & $(2.93)$     & $(1.18)$     \\ \cline{2-14} 
				& q-FedAvg \cite{li2019fair}                     & $65.20$     & $91.59$     & $44.54$       & $82.88$       & $54.87$  & \multicolumn{1}{c|}{$87.24$}  & $71.39$     & $93.88$     & $60.79$       & $90.03$       & $66.09$      & $91.95$      \\
				& (ICLR'20)                    & $(1.26)$    & $(0.67)$    & $(0.80)$      & $(0.61)$      & $(0.58)$ & \multicolumn{1}{c|}{$(0.47)$} & $(0.73)$    & $(0.30)$    & $(1.38)$      & $(0.39)$      & $(0.96)$     & $(0.30)$     \\ \cline{2-14} 
				& FairFed \cite{ezzeldin2023fairfed}                      & $60.36$     & $90.29$     & $49.04$       & $84.86$       & $54.70$  & \multicolumn{1}{c|}{$87.58$}  & $63.94$     & $91.89$     & $56.27$       & $89.58$       & $60.11$      & $90.73$      \\
				& (AAAI'23)                    & $(1.59)$    & $(0.54)$    & $(1.80)$      & $(0.64)$      & $(1.39)$ & \multicolumn{1}{c|}{$(0.52)$} & $(1.58)$    & $(0.48)$    & $(1.59)$      & $(0.58)$      & $(1.48)$     & $(0.49)$     \\ \cline{2-14} 
				& FedCE \cite{jiang2023fair}                        & $62.21$     & $90.37$     & $45.25$       & $83.37$       & $53.73$  & \multicolumn{1}{c|}{$86.87$}  & $72.29$     & $94.26$     & $61.85$       & $90.95$       & $67.07$      & $92.61$      \\
				& (CVPR'23)                    & $(1.27)$    & $(0.56)$    & $(2.65)$      & $(1.72)$      & $(1.11)$ & \multicolumn{1}{c|}{$(0.91)$} & $(0.64)$    & $(0.28)$    & $(1.28)$      & $(0.40)$      & $(0.74)$     & $(0.31)$     \\ \cline{2-14} 
				& FedGA \cite{zhang2023federated}                        & $59.56$     & $89.53$     & $48.16$       & $84.64$       & $53.86$  & \multicolumn{1}{c|}{$87.08$}  & $66.35$     & $92.78$     & $59.22$       & $90.20$       & $62.79$      & $91.49$      \\
				& (CVPR'23)                    & $(1.55)$    & $(0.55)$    & $(1.38)$      & $(0.47)$      & $(1.03)$ & \multicolumn{1}{c|}{$(0.47)$} & $(1.53)$    & $(0.45)$    & $(1.46)$      & $(0.36)$      & $(1.31)$     & $(0.34)$     \\ \hline \hline
				\multirow{2}{*}{\begin{tabular}[c]{@{}c@{}}Fair Generalization\\ (1st-Order State)\end{tabular}}     & FedISM \cite{FedISM}                   & $66.94$       & $93.21$       & $51.62$       & $86.89$       & $59.28$       & \multicolumn{1}{c|}{$90.05$}       & $71.31$       & $94.84$       & $62.54$       & $92.31$       & $66.92$       & $93.57$       \\
				& (IJCAI'24) & $(0.84)$      & $(0.32)$      & $(1.39)$      & $(0.49)$      & $(0.52)$      & \multicolumn{1}{c|}{$(0.18)$}      & $(0.51)$      & $(0.15)$      & $(0.83)$      & $(0.18)$      & $(0.36)$      & $(0.12)$      \\ \hline \hline
				\multirow{4}{*}{\begin{tabular}[c]{@{}c@{}}Ours\\ (Multiple States)\end{tabular}} & \multirow{2}{*}{FedISM+ (S)} & $\bm{69.12}$  & $\bm{94.14}$  & $50.79$       & $87.37$       & $\bm{59.96}$  & \multicolumn{1}{c|}{$\bm{90.75}$}  & $72.25$       & $95.18$       & $\bm{65.47}$  & $93.14$       & $68.86$       & $94.16$       \\
				&                              & $\bm{(1.00)}$ & $\bm{(0.26)}$ & $(0.99)$      & $(0.26)$      & $\bm{(0.81)}$ & \multicolumn{1}{c|}{$\bm{(0.15)}$} & $(0.51)$      & $(0.14)$      & $\bm{(1.28)}$ & $(0.29)$      & $(0.74)$      & $(0.19)$      \\ \cline{2-14} 
				& \multirow{2}{*}{FedISM+ (L)} & $67.24$       & $93.33$       & $\bm{52.08}$  & $\bm{88.02}$  & $59.66$       & \multicolumn{1}{c|}{$90.67$}       & $\bm{72.70}$  & $\bm{95.41}$  & $65.46$       & $\bm{93.16}$  & $\bm{69.08}$  & $\bm{94.29}$  \\
				&                              & $(1.14)$      & $(0.46)$      & $\bm{(1.66)}$ & $\bm{(0.51)}$ & $(0.89)$      & \multicolumn{1}{c|}{$(0.27)$}      & $\bm{(0.48)}$ & $\bm{(0.10)}$ & $(0.83)$      & $\bm{(0.23)}$ & $\bm{(0.43)}$ & $\bm{(0.10)}$ \\ \hline
				\bottomrule
			\end{tabular}
		}
		\label{tab:SOTA-ISIC}
	\end{table*}

	\begin{table}[!t]
		\centering
		\caption{Client-level fairness comparison. Numbers outside the parentheses correspond to the mean (\%), while those within the parentheses are the standard deviation (\%). Values in bold denote the best result, and values underlined indicate the second best.}
		\resizebox{1.0\columnwidth}{!}{
			\renewcommand{\arraystretch}{1.2}
			\begin{tabular}{l|ccccc}
				\toprule
				\hline
				\multirow{3}{*}{Method} & \multicolumn{5}{c}{STD of AUC (\%) among Clients $\downarrow$}                                                                \\ \cline{2-6} 
				& \multicolumn{2}{c}{RSNA ICH}                 & \multicolumn{2}{c|}{ISIC 2019}                               & \multirow{2}{*}{Avg} \\ \cline{2-5}
				& Gaussian Noise     & Motion Blur        & Gaussian Noise     & \multicolumn{1}{c|}{Motion Blur}        &                      \\ \hline \hline
				FedAvg \cite{mcmahan2017communication}                  & $4.08 (0.33)$      & $1.51 (0.12)$      & $5.55 (0.59)$      & \multicolumn{1}{c|}{$1.78 (0.16)$}      & $3.23$               \\ \hline \hline
				Agnostic-FL \cite{mohri2019agnostic}            & $3.28 (2.56)$      & $1.47 (1.19)$      & $5.79 (1.91)$      & \multicolumn{1}{c|}{$1.02 (0.61)$}      & $2.89$               \\
				q-FedAvg \cite{li2019fair}                & $3.02 (0.46)$      & $1.16 (0.33)$      & $3.49 (0.35)$      & \multicolumn{1}{c|}{$1.54 (0.14)$}      & $2.30$               \\
				FairFed  \cite{ezzeldin2023fairfed}                 & $2.72 (0.25)$      & $1.04 (0.15)$      & $2.17 (0.24)$      & \multicolumn{1}{c|}{$0.93 (0.17)$}      & $1.72$               \\
				FedCE \cite{jiang2023fair}                   & $2.95 (0.27)$      & $1.18 (0.11)$      & $2.80 (0.72)$      & \multicolumn{1}{c|}{$1.32 (0.12)$}      & $2.06$               \\
				FedGA \cite{zhang2023federated}                   & \uline{$2.17 (0.20)$}      & $0.79 (0.12)$      & $\bm{1.96 (0.17)}$ & \multicolumn{1}{c|}{$1.03 (0.18)$}      & \uline{$1.49$}               \\  \hline \hline
				FedISM \cite{FedISM}                 & $2.18 (0.12)$      & \uline{$0.64 (0.05)$}      & $2.53 (0.30)$      & \multicolumn{1}{c|}{\uline{$1.01 (0.10)$}}      & $1.59$               \\ \hline \hline
				FedISM+                 & $\bm{2.03 (0.11)}$ & $\bm{0.57 (0.06)}$ & \uline{$2.13 (0.32)$}      & \multicolumn{1}{c|}{$\bm{0.82 (0.10)}$} & $\bm{1.39}$          \\ \hline
				\bottomrule
			\end{tabular}
		}
		\label{tab:fairness}
	\end{table}

	\subsubsection{Classification Performance Comparison}
	
	Quantitative comparison results measured by the mean and standard deviation (stemming from multiple runs) are summarized in Tabs. \ref{tab:SOTA-ICH} and \ref{tab:SOTA-ISIC}. Under the given quality shifts, FedAvg demonstrates a clear performance bias toward clean images, resulting in a diminished lower performance bound across distributions. To address this, previous fair optimization methods focus primarily on aligning the 0th-order state of convergence. In this way, most methods improve the performance on corrupted images, with the exception of the less stable Agnostic-FL. As discussed before, they are sub-optimal due to their neglect of the 1st-order state. In contrast, FedISM, which focuses on the sharpness of the loss surface, achieves better generalization on corrupted images. For instance, on the Gaussian noise-corrupted distribution of the RSNA ICH dataset, FedISM outperforms FedAvg and the best fair optimization method (\textit{i.e.}, FedGA) by 10.85\% and 3.78\% in ACC, respectively. However, FedISM’s limited perception of the loss surface restricts its generalization capacity estimation, thus impairing performance. This limitation is addressed by FedISM+, which progressively increases the search distance from zero, considering both 0th- and multiple 1st-order states. This approach better bridges the gap between training and testing distributions across different clients. The results demonstrate that FedISM+ surpasses FedISM in most metrics for RSNA ICH and all metrics for ISIC 2019; for instance, on the motion blur-corrupted distribution of the ISIC 2019 dataset, FedISM+ outperforms FedISM by more than 0.70\% in AUC, exceeding the sum of standard deviations. These outcomes highlight the superiority of our design.

	\subsubsection{Client-Level Fairness Comparison}
	Rather than focusing solely on the worst-performing distributions, fairness in FL can also be evaluated by the uniformity of classification performance among clients \cite{li2019fair}. To evaluate this, we report the standard deviation of performance among clients across multiple runs in Tab. \ref{tab:fairness}. In our experiments, we simulate client-level testing by distributing the entire clean/corrupted testing set to specific clients. The results show that our proposed solution, FedISM+, achieves superior uniformity. While FedGA \cite{zhang2023federated} also performs well, even outperforming FedISM+ in one case, it primarily achieves this uniformity by suppressing performance on clean images (see Tabs. \ref{tab:SOTA-ICH} and \ref{tab:SOTA-ISIC}). Such an approach, though it may produce uniform performance, is less meaningful in practice because it does not truly enhance the model’s ability to handle diverse distributions.
	
	\textbf{Remark:} Compared to existing solutions, our proposed method, FedISM+, not only enhances classification performance on the most challenging distributions but also improves the uniformity of performance across all clients.

	\subsection{Discussion}
	\subsubsection{Discussion on Search Distance}
	In this section, we empirically evaluate our progressive scheme of search distance $\rho$ from 0 to $\rho_{max}$ (set as 0.1 in experiments) using the experimental setting mentioned in Sec. \ref{sec:component_ablation}.
	
\begin{figure*}[!t] 
		\centering
		\includegraphics[width=1.0\textwidth]{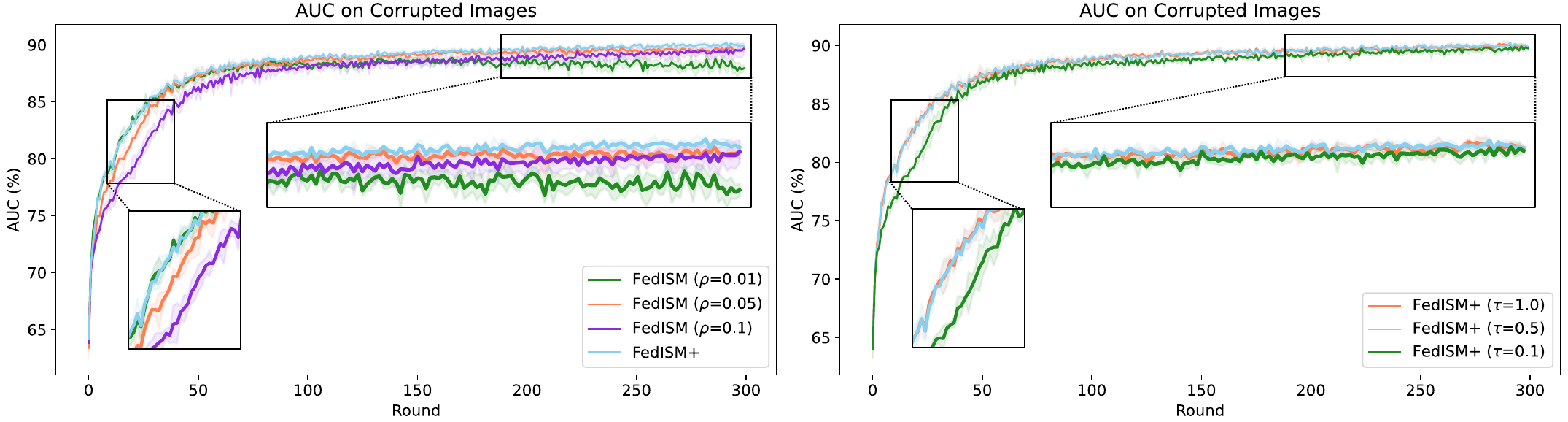}
		\caption{\textbf{Left}: Comparison of FedISM+ with FedISM across different $\rho$ values. \textbf{Right}: Evaluation across different $\tau$ settings. The solid lines indicate the mean values, and the shaded regions represent the standard deviations.}
		\label{fig:rho}
	\end{figure*}

	Firstly, we show that this scheme is not equivalent to simply tuning $\rho$ for FedISM \cite{FedISM} that considers sharpness defined at a fixed search distance. As shown in the left of Fig. \ref{fig:rho}, FedISM shows a slower convergence rate with the increase of $\rho$ in the early stage. This is because larger $\rho$ can lead to more focus on the 1st-order state and slow down the decrease of loss. However, in the last stage, larger $\rho$ helps the model converge on a flat minimum, which is better for generalization and achieves high classification performance on the testing set. Our method FedISM+, by increasing $\rho$ as training progresses and considering both 0th- and 1st-order states, achieves a quicker convergence rate in the beginning and better performance in the later stage.
	
	Secondly, we validate the effectiveness of $\tau$, which controls the rate at which $\rho$ evolves, as illustrated on the right side of Fig. \ref{fig:rho}. By decreasing $\tau$, $\rho$ reaches its maximum more quickly, exhibiting a trend similar to FedISM with a large $\rho$ in the early stages. Because FedISM+ considers both 0th- and 1st-order states, it achieves relatively good convergence in the later stages, even with $\tau$ tuning.
	
	\textbf{Remark:} Our proposed dynamic $\rho$ strategy is not trivial as its effectiveness can not be achieved by simply tuning $\rho$ in FedISM. The increase rate of $\rho$ can impact the convergence rate of FedISM+.
	
	\begin{figure}[!t] 
		\centering
		\includegraphics[width=1.0\columnwidth]{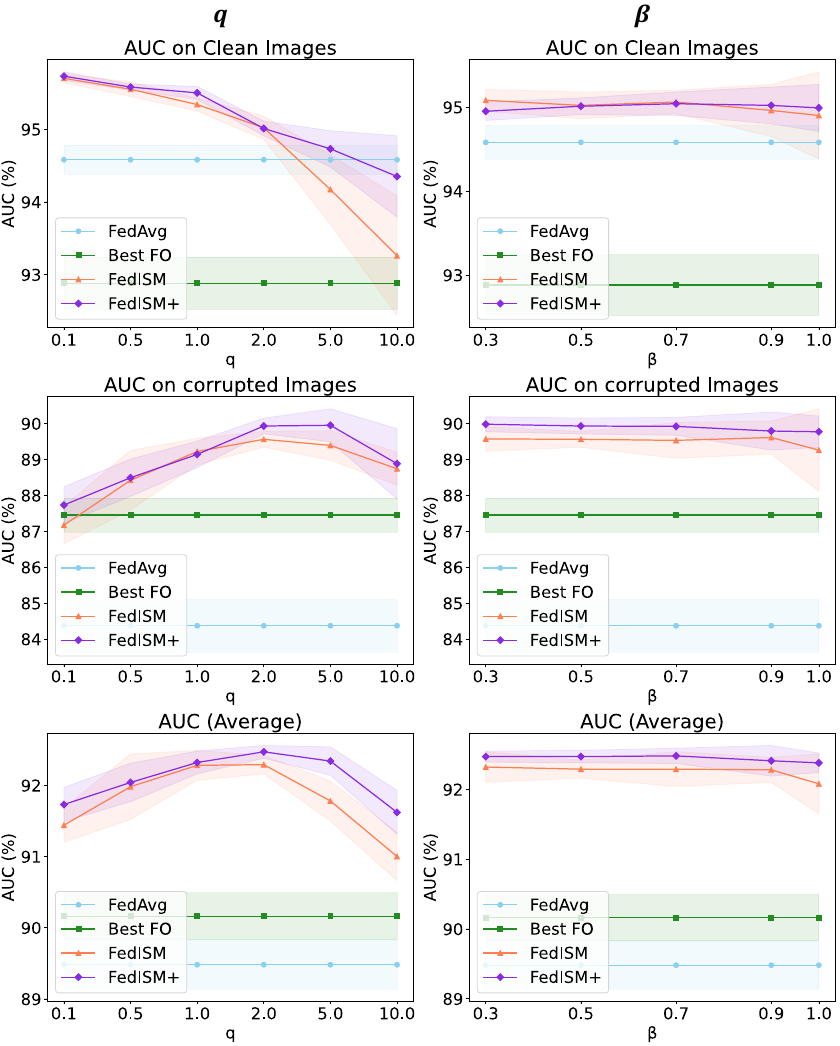}
		\caption{\textbf{Left}: Evaluation on different $q$. \textbf{Right}: Evaluation on different $\beta$. The solid lines indicate the mean values, and the shaded regions represent the standard deviations. The best fair optimization method from Section \ref{sec:sota} is denoted as ``Best FO'' according to the performance on corrupted images.}
		\label{fig:q&beta}
	\end{figure}

	\subsubsection{Discussion on Parameter $q$} \label{sec:discussion_q}
	
	\begin{figure}[!t] 
		\centering
		\includegraphics[width=1.0\columnwidth]{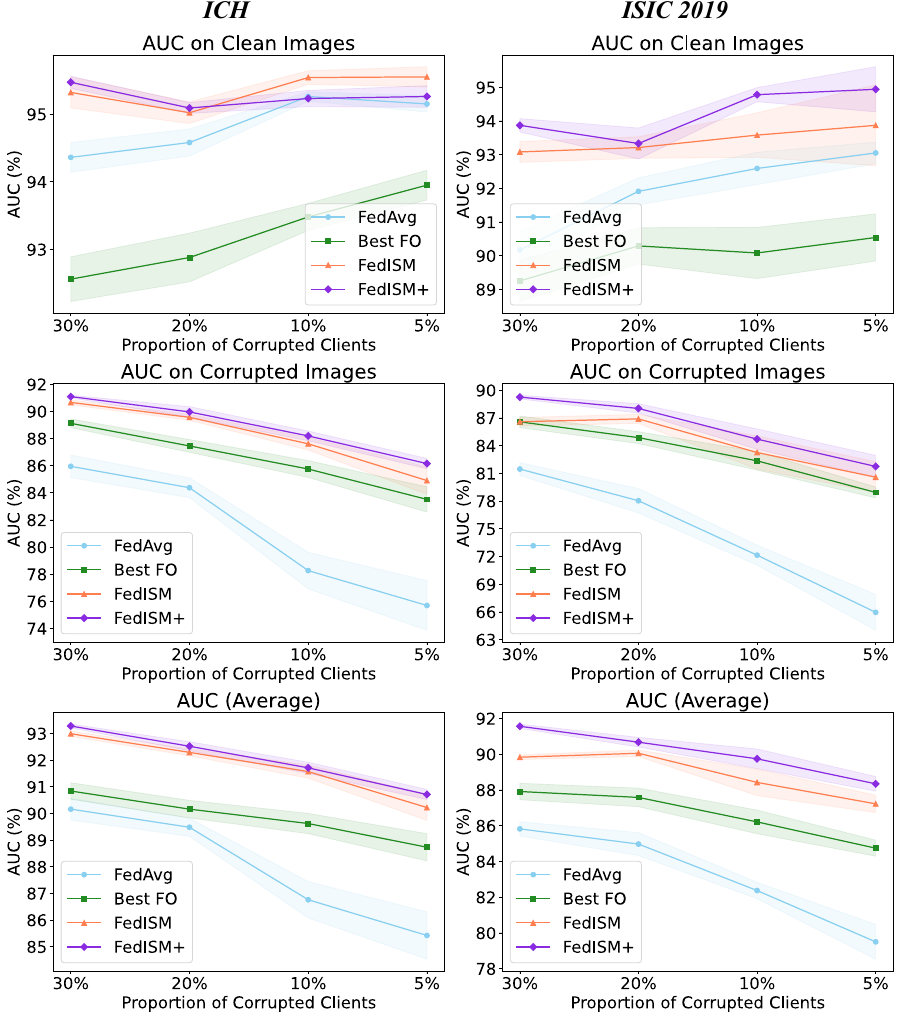}
		\caption{Evaluation with varying corruption ratios (\textit{e.g.}, 30\% indicates a 3:7 ratio of corrupted to clean clients), with the left column representing experiments on RSNA ICH and the right column on ISIC 2019. The solid lines indicate the mean values, and the shaded regions represent the standard deviations. The best fair optimization method from Section \ref{sec:sota} is denoted as ``Best FO'' according to the performance on corrupted images. We tune $\tau$ from \{0.1, 0.5, 1.0\}.}
		\label{fig:robustness}
	\end{figure}
	
	The parameter $q$ plays a key role in determining how much emphasis FedISM+ places on clients with higher sharpness. To investigate its effect, we vary $q$ over the values \{0.1, 0.5, 1.0, 2.0, 5.0, 10.0\} and evaluate the performance using the setup described in Sec. \ref{sec:component_ablation}, with results shown in the left column of Fig. \ref{fig:q&beta}. For reference, we also report the performance of FedAvg \cite{mcmahan2017communication}, the best fair optimization method discussed in Sec. \ref{sec:sota} and FedISM \cite{FedISM}. Both FedISM and FedISM+ exhibit similar trends as $q$ varies. In particular, increasing $q$ causes both methods to focus more on clients with higher sharpness (\textit{i.e.}, clients with corrupted images), which leads to improved performance on corrupted images but a slight decline on clean images. Notably, very large values of $q$ (such as 5 and 10) result in a minor performance drop on corrupted images, likely due to instability in the training process. On the whole, FedISM(+) performs well across a wide range of $q$ values, alleviating the need for extensive parameter tuning in practical applications. Additionally, FedISM+ generally outperforms FedISM, highlighting the effectiveness of the new design introduced in this paper.

	\subsubsection{Discussion on Parameter $\beta$}
	To maintain stability during training, we utilize a moving average scheme to process the aggregation weights in FedISM+ as follows:
	\begin{equation} \nonumber
		{\boldsymbol w}_{t} = \beta \widetilde{\boldsymbol w}_{t} + (1-\beta) {\boldsymbol w}_{t-1}.
	\end{equation}
	In this section, we investigate the influence of the parameter $\beta$ by experimenting with different values \{0.3, 0.5, 0.7, 0.9, 1.0\}, using the setup described in Sec. \ref{sec:component_ablation}. The results of these experiments are presented in the right column of Fig. \ref{fig:q&beta}, alongside the performance of FedAvg \cite{mcmahan2017communication}, the best fair optimization method, and FedISM \cite{FedISM} for a more thorough comparison. The figures show that, compared to cases where a moving average is either not applied or only weakly applied (\textit{i.e.}, $\beta=1.0$ or $0.9$), incorporating a moderate moving average leads to better and more stable performance, often with a reduced standard deviation. Furthermore, the performance tends to be relatively unaffected by variations in $\beta$ when the moving average is used.
	
	\subsubsection{Robustness across Different Shift Degrees}
	Our experiments also demonstrate the robustness of FedISM+, showcasing stable performance across various degrees of shift, specifically different ratios of corrupted clients. Quantitative results for two datasets at varying proportions (\textit{i.e.}, 30\%, 20\%, 10\% and 5\%) of clients with Gaussian noise-corrupted images are illustrated in Fig. \ref{fig:robustness}. The performance of FedAvg \cite{mcmahan2017communication} reveals that reducing the ratio of corrupted clients complicates the maintenance of performance on corrupted images. FedISM consistently surpasses both FedAvg and previous fair optimization methods across all ratios. Notably, on both datasets, both the AUC on corrupted images and the average AUC for FedISM+ surpass those of FedISM, demonstrating the effectiveness of our new design.
	
	\begin{figure}[!t] 
		\centering
		\includegraphics[width=1.0\columnwidth]{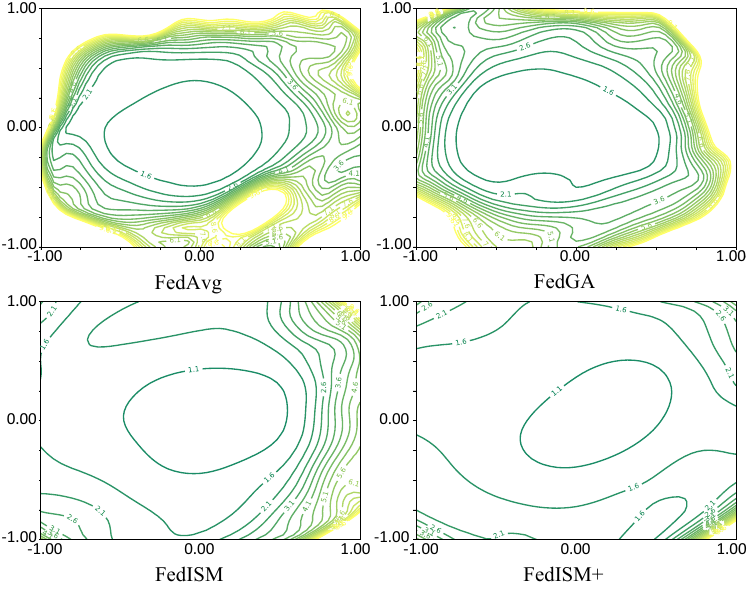}
		\caption{Visualization of loss landscape on corrupted testing sets.}
		\label{fig:loss-landscape}
	\end{figure}
	
	\begin{table}[!t]
		\centering
		\caption{Communication overhead of each client in each round. 44.59MB is the size of ResNet-18 \cite{he2016deep}.}
		\renewcommand{\arraystretch}{1.0}
		\begin{tabular}{l|cc}
			\toprule
			\hline
			Method                 & FedAvg \cite{mcmahan2017communication}  & FedISM+                \\ \hline
			Overhead  & 2$\times$44.59MB & 2$\times$44.59MB+4B (1.000000043$\times$) \\ \hline
			\bottomrule
		\end{tabular}
		\label{tab:communication}
	\end{table}
	
	\subsubsection{Visualization of Loss landscape}
	Following \cite{li2018visualizing}, we visualize the loss landscape under model weight perturbation in Fig. \ref{fig:loss-landscape}. The experimental settings are the same as in Sec. \ref{sec:component_ablation}. Compared with FedAvg \cite{mcmahan2017communication} and FedGA \cite{zhang2023federated}, FedISM(+) converges to flatter regions, aligning with our aims and motivations. Notably, FedISM+ achieves smaller sharpness over a wider region compared to FedISM, as it considers more states.
	
	\subsubsection{Communication Overhead and Privacy} \label{sec:privacy}
	FedISM+ uploads sharpness/perturbed loss to the server, which can raise concerns regarding communication overhead and privacy. These concerns are addressed in this section. Regarding the former, we demonstrate that the additional communication overhead of FedISM+ is negligible, as shown in Tab. \ref{tab:communication}, since only one extra float value needs to be uploaded. Concerning the latter, existing techniques in secure multi-party computation (\textit{e.g}., Homomorphic Encryption) ensure that we can compute the sum of all sharpness/perturbed loss (Eq. \ref{eq:FedISM_weights} and Eq. \ref{eq:FedISM+_weights}) without leaking any individual value \cite{shen2022agnostic}. This prevents third parties from inferring any information from the uploaded values.

	\section{Conclusion}
	In this paper, we explore how to address the fairness problem of FL raised by imaging quality shifts. Instead of promoting fairness through aligning a 0th- or 1st-order state of convergence across clients, we first propose capturing the full spectrum of convergence to build a better surrogate of fairness. Specifically, we generalize previous approaches by considering multiple states, and calculating sharpness/perturbed loss at different distances from zero up to a maximum. Building on this concept, we propose FedISM+, which effectively aligns the model’s states across different clients. Comprehensive experiments conducted on the widely recognized RSNA ICH and ISIC 2019 datasets show that FedISM+ outperforms existing state-of-the-art fair FL approaches. We believe that our findings will inspire more studies on fair FL in medical applications and beyond, and generalization estimation with metrics on the training set.

	%\begin{thebibliography}{1}
	
	\bibliographystyle{IEEEtran}
	\bibliography{IEEEabrv,Bibliography}

\end{document}